\documentclass[superscriptaddress,preprintnumbers,prb,aps,twocolumn, nofootinbib]{revtex4-2}

\usepackage{amsfonts}
\usepackage{amssymb}
\usepackage{amsmath}
\usepackage{graphicx}
\usepackage{natbib}
\usepackage{physics}
\usepackage{siunitx}
\usepackage{float}
\usepackage{multirow}
\usepackage{ulem}
\usepackage{url}

\usepackage{fancyhdr}

\setcitestyle{numbers,square}

\usepackage[usenames,dvipsnames]{color}
\usepackage{soul}

\usepackage[hidelinks]{hyperref} 

\usepackage{tablefootnote}
\makeatletter
\newcommand\footnoteref[1]{\protected@xdef\@thefnmark{\ref{#1}}\@footnotemark}
\makeatother

 \makeatletter
\def\@fnsymbol#1{\ensuremath{\ifcase#1\or \dagger\or \ddagger\or
   \mathsection\or \mathparagraph\or \|\or **\or \dagger\dagger
   \or \ddagger\ddagger \else\@ctrerr\fi}}
    \makeatother

\usepackage{fancyhdr}            % Permits header customization. See header section below.
\fancypagestyle{plain}{
    \lhead{}
    \fancyhead[R]{\thepage}
    \fancyhead[L]{}
    
    \fancyfoot{}
}

\begin{document}

\newcommand{\ie}{{\it i.e.}}
\newcommand{\eg}{{\it e.g.}}
\newcommand{\etal}{{\it et al.}}

\newcommand{\TVO}{TmVO$_{4}$}
\newcommand{\YVO}{YVO$_{4}$}
\newcommand{\YTVO}{Tm$_{0.7}$Y$_{0.3}$VO$_{4}$}

\newcommand{\micron}{$\mu$m}

\newcommand{\Kzy}{$\kappa_{\rm {zy}}$}
\newcommand{\Kzz}{$\kappa_{\rm {zz}}$}

\newcommand{\ncco}{Nd$_{2-x}$Ce$_x$CuO$_4$}
\newcommand{\pcco}{Pr$_{2-x}$Ce$_x$CuO$_4$}
\newcommand{\ndlsco}{La$_{1.6-x}$Nd$_{0.4}$Sr$_x$CuO$_4$}

\newcommand{\TN}{$T_{\rm {N}}$}
\newcommand{\Tc}{$T_{\rm {c}}$}

\title{Role of magnetic ions in the thermal Hall effect of the paramagnetic insulator TmVO$_{4}$}

\author{Ashvini Vallipuram$^{\star}$}
\thanks{$^{\star}$ A.V. and L.C. contributed equally to this work.}
\email{ashvini.vallipuram@usherbrooke.ca}
\affiliation{Institut quantique, D\'epartement de physique \& RQMP, Universit\'e de Sherbrooke, Sherbrooke, Qu\'ebec, Canada J1K 2R1}

\author{Lu Chen$^{\star}$}
\email{lu.chen@usherbrooke.ca}
\affiliation{Institut quantique, D\'epartement de physique \& RQMP, Universit\'e de Sherbrooke, Sherbrooke, Qu\'ebec, Canada J1K 2R1}

\author{Emma Campillo}
\affiliation{Institut quantique, D\'epartement de physique \& RQMP, Universit\'e de Sherbrooke, Sherbrooke, Qu\'ebec, Canada J1K 2R1}

\author{Manel Mezidi}
\affiliation{Institut quantique, D\'epartement de physique \& RQMP, Universit\'e de Sherbrooke, Sherbrooke, Qu\'ebec, Canada J1K 2R1}
\affiliation{Université Paris-Cité, Laboratoire Matériaux et Phénomènes Quantiques, CNRS (UMR 7162), 75013 Paris, France}

\author{Gaël Grissonnanche}
\affiliation{Laboratory of Atomic and Solid State Physics, Cornell University, Ithaca, New York 14853, USA}
\affiliation{Kavli Institute at Cornell for Nanoscale Science, Ithaca, New York 14853, USA}

\author{Marie-Eve Boulanger}
\affiliation{Institut quantique, D\'epartement de physique \& RQMP, Universit\'e de Sherbrooke, Sherbrooke, Qu\'ebec, Canada J1K 2R1}

\author{Étienne Lefrançois}
\affiliation{Institut quantique, D\'epartement de physique \& RQMP, Universit\'e de Sherbrooke, Sherbrooke, Qu\'ebec, Canada J1K 2R1}

\author{Mark P. Zic}
\affiliation{Geballe Laboratory for Advanced Materials and Department of Physics, Stanford University, California 94305, USA}

\author{Yuntian Li}
\affiliation{Geballe Laboratory for Advanced Materials and Department of Applied Physics, Stanford University, California 94305, USA}

\author{Ian R. Fisher}
\affiliation{Geballe Laboratory for Advanced Materials and Department of Applied Physics, Stanford University, California 94305, USA}

\author{Jordan Baglo}
\affiliation{Institut quantique, D\'epartement de physique \& RQMP, Universit\'e de Sherbrooke, Sherbrooke, Qu\'ebec, Canada J1K 2R1}

\author{Louis~Taillefer}
\email{louis.taillefer@usherbrooke.ca}
\affiliation{Institut quantique, D\'epartement de physique \& RQMP, Universit\'e de Sherbrooke, Sherbrooke, Qu\'ebec, Canada J1K 2R1}
\affiliation{Canadian Institute for Advanced Research, Toronto, Ontario, Canada M5G 1M1}

\date{\today}

\begin{abstract}

In a growing number of materials, phonons have been found to generate a thermal Hall effect, but the underlying mechanism remains unclear. Inspired by previous studies that revealed the importance of Tb$^{3+}$ ions in generating the thermal Hall effect in a family of pyrochlores, 
we investigated the role of Tm$^{3+}$ ions in \TVO, a paramagnetic insulator with a different crystal structure.
We observe a negative thermal Hall conductivity in \TVO \hspace{0.1em} with a magnitude such that the Hall angle, $|\kappa_{xy}$/$\kappa_{xx}|$, 
is approximately 1 $\times$ 10$^{-3}$ at $H$ = 15 T and $T$ = 20 K, typical for a phonon-generated thermal Hall effect. In contrast to the negligible $\kappa_{xy}$ 
found in the nonmagnetic pyrochlore analog (where the Tb$^{3+}$ ions are replaced with Y$^{3+}$), we observe a negative $\kappa_{xy}$ in \YVO \hspace{0.1em} with a Hall angle of magnitude comparable to that of \TVO. 
This shows that the Tm$^{3+}$ ions are not essential for the thermal Hall effect in this family of materials. 
Interestingly, at an intermediate Y concentration of $x$ = 0.3 in Tm$_{1-x}$Y$_{x}$VO$_{4}$, 
$\kappa_{xy}$ was found to have a positive sign, pointing to the importance of impurities in the thermal Hall effect of phonons.
\end{abstract}

\pacs{Valid PACS appear here}% PACS, the Physics and Astronomy


\maketitle

\section{INTRODUCTION}

%Thermal transport measurements have always been a powerful tool to gain information about materials that do not conduct electricity. It is quite useful when wanting to detect exotic excitations. However, i

\begin{figure*}[t]
    \centering
    \includegraphics[width=0.8\textwidth]{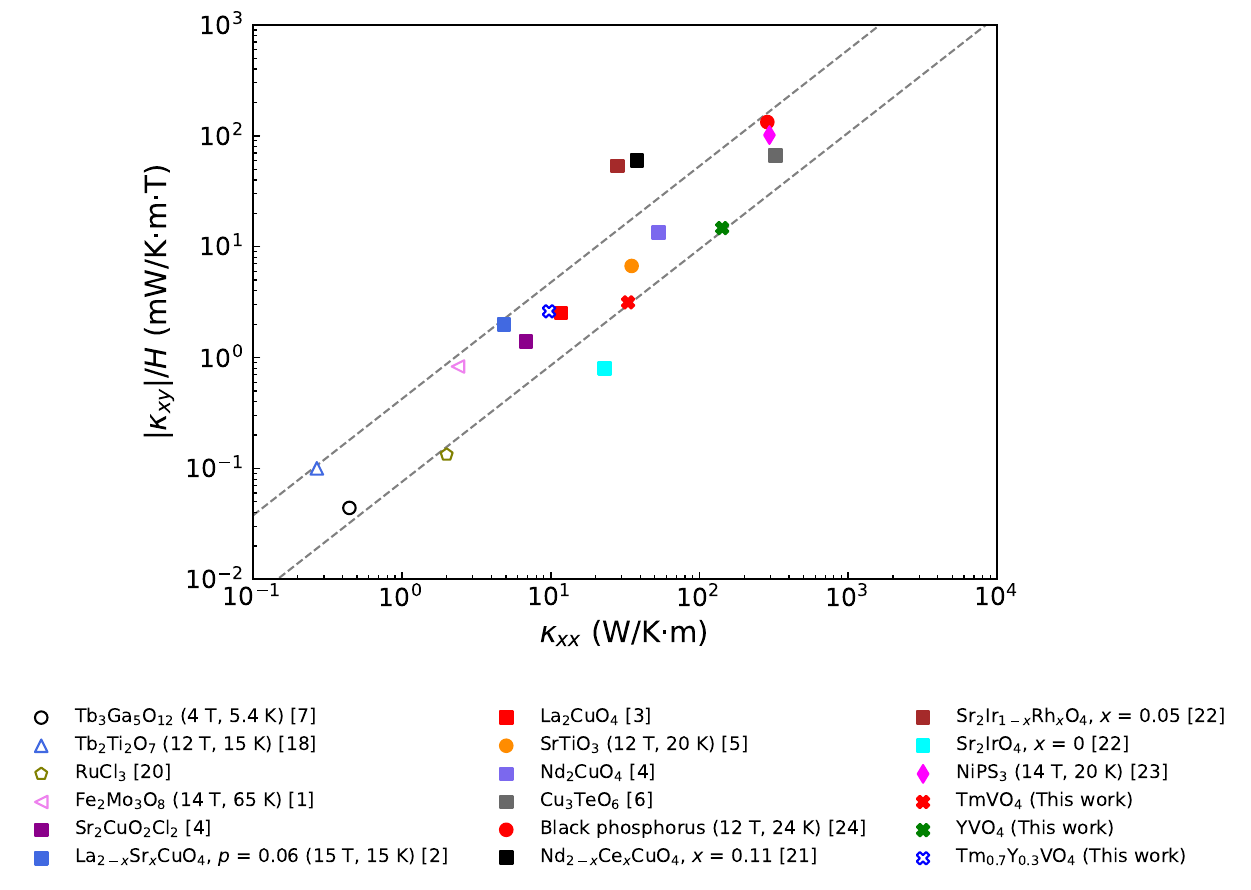}
    \caption{Phonon thermal Hall conductivity normalized by field, $|\kappa_{xy}|/H$ of various insulators \cite{Tb3Ga5O12_Kxy_Strohm_2005, Phononic_thermal_Hall_effect_in_diluted_terbium_oxides, THE_Multiferroics, Grissonnanche_2020_Chiral_phonons, Boulanger_2020_Kxy_NCO_SCOC, Sr2TiO4_Kamran_2020, CTO_Lu_2022, RuCL3_Etienne_2022, Boulanger_2022_Kxy_electron_dopes_cuprates, ataei2023impurityinduced, NiPS3} as a function of their thermal conductivity, $\kappa_{xx}$. The data are taken at $H$ = 15 T and $T$ = 20 K unless indicated otherwise. This figure is inspired by a similar plot from Li \textit{et al.} \cite{Kamran_2023_Black_phosphorous}, to which other data points have been added, including those from our three samples. The gray lines mark the region where most values of $\kappa_{xy}$ are found, showing that $\kappa_{xy}$ scales with $\kappa_{xx}$, as emphasized previously \cite{THE_Multiferroics, Boulanger_2020_Kxy_NCO_SCOC, CTO_Lu_2022, Kamran_2023_Black_phosphorous}. The data points that lie outside the delineated region are Nd$_{2-x}$Ce$_{x}$CuO$_{4}$ at $x$ = 0.11 \cite{Boulanger_2022_Kxy_electron_dopes_cuprates} and Sr$_{2}$Ir$_{1-x}$Rh$_{x}$O$_{4}$ at $x$ = 0 as well as $x$ = 0.05 \cite{ataei2023impurityinduced} (see text). Open (full) symbols indicate a positive (negative) $\kappa_{xy}$.}
    
    \label{fig:ratio_pthe}
\end{figure*}

In the last decade, a 
%non-zero 
thermal Hall effect has been measured in a number of insulators where phonons are the main heat carriers,
including 
multiferroics such as Fe$_2$Mo$_3$O$_8$~\cite{THE_Multiferroics},
cuprates such as La$_{2}$CuO$_{4}$~\cite{2019_Giant_Kxy_pseudogap_cuprates,Grissonnanche_2020_Chiral_phonons}
and Nd$_{2}$CuO$_{4}$~\cite{Boulanger_2020_Kxy_NCO_SCOC},
nonmagnetic SrTiO$_{3}$~\cite{Sr2TiO4_Kamran_2020},
and
the antiferromagnet Cu$_{3}$TeO$_{6}$~\cite{CTO_Lu_2022}, see Fig. \ref{fig:ratio_pthe}.
%
%Hence, this has puzzled a lot of researchers leading to a wide range of studies on materials such as Tb$_{2}$Ti$_{2}$O$_{7}$,  SrTiO$_{3}$, Sr$_{2}$CuO$_{2}$Cl$_{2}$, Cu$_{3}$TeO$_{6}$, RuCl$_{3}$ and black phosphorus \cite{Hirschberger_2015_pyrochlores, THE_Multiferroics, Phononic_thermal_Hall_effect_in_diluted_terbium_oxides, 2019_Giant_Kxy_pseudogap_cuprates, Sr2TiO4_Kamran_2020, Boulanger_2020_Kxy_NCO_SCOC, Grissonnanche_2020_Chiral_phonons, Boulanger_2022_Kxy_electron_dopes_cuprates, CTO_Lu_2022, RuCL3_Etienne_2022, ataei2023impurityinduced, Kamran_2023_Black_phosphorous}. 
%
In Tb$_{3}$Ga$_{5}$O$_{12}$~\cite{Tb3Ga5O12_Kxy_Strohm_2005},
the first insulator in which a thermal Hall signal was detected,
the effect was attributed to skew scattering of phonons by superstoichiometric Tb$^{3+}$ impurities~\cite{Mori_2014}. 

%was the first material to ever show such a surprising result. The measured transverse temperature difference is understood to be because of the coupling of certain phonon modes to the applied $B$. Hence, anisotropic scattering can be generated in this material and phonons scatter off of the superstochiometric Tb$^{+3}$ ions (resonant skew scattering), which are impurities, to give a non-zero K$_{xy}$ \cite{Mori_2014}. \\

Since then, 
a number of theoretical scenarios have been proposed to explain the origin of the phonon thermal Hall effect.
%
 %trying to explain this new physics based on all the recent experimental developments. 
Some attribute the thermal Hall conductivity $\kappa_{xy}$~to the Berry curvature of phonon bands \cite{Berry_curvature_PTHE}. 
Others link it to various types of spin-lattice coupling~\cite{ye2021phonon,Zhang_ph_magnon_coupling,Samajdar_ph_magnon_theory,Mangeolle_2022}. 
In yet others, the role of impurities is considered important~\cite{Guo_point_like, 
Flebus_PTHE,XQ_Sun_resonant_skew_scatte,Guo_Joshi_Sachdev_2022}, but it remains unclear which of these scenarios applies to what material.

In a previous study on pyrochlores~\cite{Hirschberger_2015_pyrochlores}, 
a sizable $\kappa_{xy}$~was observed in Tb$_{2}$Ti$_{2}$O$_{7}$ but a negligible $\kappa_{xy}$~was detected in Y$_{2}$Ti$_{2}$O$_{7}$.
Although the $\kappa_{xy}$~signal in Tb$_{2}$Ti$_{2}$O$_{7}$ was originally attributed to some exotic neutral excitations linked to the spin-liquid nature of the system~\cite{Hirschberger_2015_pyrochlores},
it was later argued that phonons are, in fact, responsible for the thermal Hall effect in this material~\cite{Phononic_thermal_Hall_effect_in_diluted_terbium_oxides},
since a $\kappa_{xy}$~signal of comparable magnitude is still observed when 70\% of the Tb$^{3+}$ ions are replaced by Y$^{3+}$ ions,
and the spin state of the system is profoundly altered.
Irrespective of the underlying mechanism, this study does show that Tb$^{3+}$ ions are essential for generating a thermal Hall effect in these pyrochlores.

%It is thus important to understand the origin of this phenomena. 
%Hence, similar to a previous study on pyrochlores (Tb$_{2}$Ti$_{2}$O$_{7}$ and Y$_{2}$Ti$_{2}$O$_{7}$), we want to investigate the role of the Tm$^{3+}$ magnetic ions in the paramagnetic insulator \TVO. In the pyrochlores study, it was found that without the magnetic Tb ions, a very negligible, almost 0, thermal Hall effect was measured in Y$_{2}$Ti$_{2}$O$_{7}$ see panel (b) of Fig. \ref{fig:Sun_Ong_pyro}. Despite the Hall angle, $\kappa_{xy}$/$\kappa_{xx}$ being very close to 0, the peak value of $\kappa_{xx}$ is 4 times bigger for Y$_{2}$Ti$_{2}$O$_{7}$ see panel (a) of Fig. \ref{fig:Sun_Ong_pyro}. \\

Here we report a similar study carried out on a different oxide, \TVO,
wherein we investigate what happens to the thermal Hall effect when we substitute the magnetic ion Tm$^{3+}$ for the nonmagnetic ion Y$^{3+}$.

The insulator \TVO \hspace{0.1em} exhibits a rich set of phenomena at low temperatures.
%
%has interesting low temperature physics that has been the subject of investigation since the late 1960s. 
%
It undergoes a cooperative Jahn-Teller phase transition at $T =T_{\mathrm{D}} \simeq 2$~K in zero field, 
and at $H = H_{\mathrm{D}} \simeq 0.5$~T at $T = 0$ (for $H // c$) \cite{vinograd_shirer_massat_2022}. (Note that we use $H$ as a shorthand notation for $\mu_{0}H$.)
The ordered state consists of a simultaneous ferroquadrupole order of the local 4$f$ electronic orbitals of each Tm atom, accompanied by a structural transition, by which the crystal structure goes from tetragonal at high temperature ($I$41/$amd$ space group [21]) to orthorhombic at low temperature \cite{PhysRevB.104.205137, Field_tuned_FQ, vinograd_shirer_massat_2022, nian2023spinecho, zic2023giant}.
%
%Since at least one Tm$^{3+}$ ion per unit cell is needed for this structural transition to occur \cite{Gehring_Gehring_1975}, no transitions are expected for \YVO \hspace{0.1em} and \YTVO. 
%
However, no magnetically ordered state has been reported in \TVO \hspace{0.1em} \cite{vinograd_shirer_massat_2022}. 
Our own study will focus on temperatures above $T_{\mathrm{D}}$.

%As for magnetic susceptibility measurements, the c axis corresponds to the easy axis \cite{Suzuki_Inoue_Ohtsuka_1981}. A large anisotropy in the g-factor is reported, where $g_{c}$ = 10 while $g_{a} = g_{b}$ = 0 \cite{Cooke_Swithenby_Wells_1972}. Hence, there exists a big magnetic torque along the c axis. \\

%In this study, we want to investigate the "high temperature" physics of this material, something that has not been done a lot in these rare-earth vanadates. Here, we want to know what happens to $\kappa_{xy}$ when the Tm (magnetic ion) in \TVO \hspace{0.1em} is substituted with Y (non-magnetic ion) where Y$^{3+}$ has a much smaller magnetic moment than Tm$^{3+}$ (4$\mu_{B}$ \cite{mom_Tm} and 1 $\mu_{B}$ \cite{mom_Y} respectively). The idea is to decrease the possible spin-phonon coupling, since the magnetic moment of the crystal is weakened, and see how that affects $\kappa_{xy}$.\\

\section{METHODS}

\subsection{Samples}
The following three rare-earth vanadates were studied: \TVO, \YVO, and \YTVO. 
The samples were grown 
%at Stanford by the group of Ian Fisher 
using a flux growth method \cite{Feigelson1968-Growth, Smith1974-Growth}.
They are needle-shaped with their long axis along the $c$ axis, 
which is the easy axis of magnetization~\cite{Suzuki_Inoue_Ohtsuka_1981}.
%
%As for magnetic susceptibility measurements, the c axis corresponds to the easy axis \cite{Suzuki_Inoue_Ohtsuka_1981}. 
%A large anisotropy in the g-factor is reported, where $g_{c}$ = 10 while $g_{a} = g_{b}$ = 0 \cite{Cooke_Swithenby_Wells_1972}. 
%Hence, there exists a big magnetic torque along the c axis. \\
%
Because \TVO~has a large anisotropy in its $g$-factor ($g_{c}$ = 10 and $g_{a} = g_{b}$ = 0)~\cite{Cooke_Swithenby_Wells_1972}, 
even a slight misalignment of the magnetic field can cause a large torque on the sample, which can detach it from its mount.

Two measures were adopted to prevent this.
First, 
the thermal transport contacts on \TVO~and \YTVO~samples were made with silver epoxy to ensure strong contacts that would adhere to a very smooth surface and would not peel off. 
These contacts were solidified for about 10 minutes on a hot plate at 150$^{\circ}$~C. 
Note that the contacts on the \YVO~samples were done with silver paint, since there is no magnetic torque in this case. 
For the heater contact, a 50-$\mu$m diameter silver wire was used and for the thermal transport contacts, wires with a diameter of 25 $\mu$m were used.

In addition, a small wood stick was positioned along the \TVO~and \YTVO~samples,
and was linked to the sample by the heater wire.
%This wood stick is connected to the sample in parallel through the heater contact and some of the heat goes through it. 
%
The thermal conductivity $\kappa_{xx}$~was measured (in zero field) with and without the wood stick and only a 5 \% (8 \%) difference 
was observed at the peak temperature for \TVO~(\YTVO). 
%
%The samples were all mounted on a copper heat sink which acts as a heat sink.
The sample dimensions (contact dimensions) are given in Table \ref{tab:geo_fact}. 

\subsection{Thermal transport measurements}
\begin{figure}[t]
    \centering
    \includegraphics[width=0.4\textwidth]{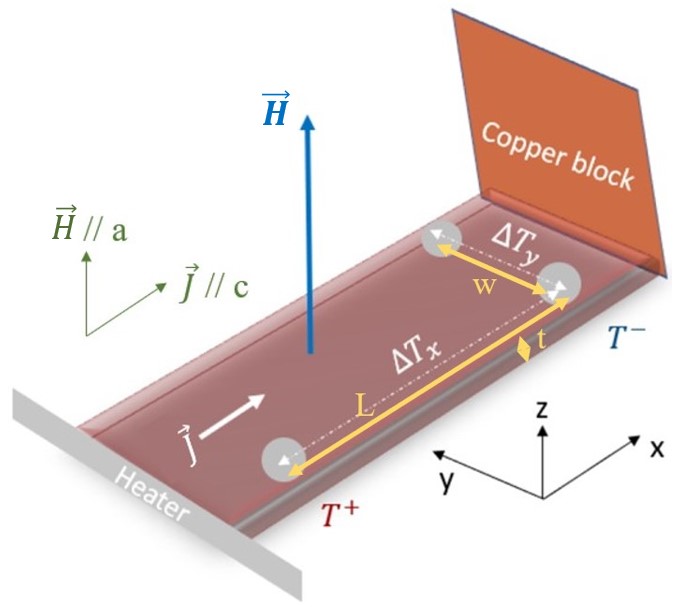}
    \caption{
  Experimental setup used to measure $\kappa_{xx}$ and $\kappa_{xy}$. 
  A heat current $J$ is generated along the $x$ axis ($c$ axis of the sample) and a magnetic field $H$ is applied perpendicular to it, along the $z$ axis ($a$ axis of the sample). 
  Differences in temperature are measured along the $x$ axis (longitudinal temperature difference, $\Delta T_{x} = T^{+} - T^{-}$) and the $y$ axis (transverse temperature difference, $\Delta T_{y}$).}
    \label{fig:setup}
\end{figure}

To perform a conventional thermal transport study, 
a five-contact measurement is done using a steady-state method at a fixed field, as sketched in Fig.~\ref{fig:setup}. 
A heat current ($\vec{J}$) is generated along the $x$ axis from the heater contact on one end of the sample and the copper block on the other end. 
The heater is a strain gauge of 5 k$\Omega$ whose resistance does not change with temperature or field. (No potential strain is conferred to the sample since the heater is glued with GE varnish to a silver pad.)
%making it a reliable heat source for thermal transport measurements. 
This heat current was applied along the $c$ axis of the sample (the $x$ axis in Fig.~\ref{fig:setup}), 
while the field $H$ was applied along the $a$ axis (the $z$ axis in Fig.~\ref{fig:setup}).
Note that the field ($H$ = 10 or 15 T) was applied at high temperatures (at 83 K) and the samples were cooled down to $\simeq 3$~K using a variable temperature insert (VTI). This procedure ensures that no hysteresis due to the sample is generated.

\begin{table}[b]
  \centering
    \begin{tabular}{ |p{3cm}||p{1.5cm}|p{1.5cm}|p{1.5cm}|  } 
        \hline
        & $L$ (mm) & $w$ (mm)  & $t$ (mm)\\ \hline
        \TVO   & 0.66    & 0.24 & 0.04\\
        \YVO&   0.49  & 0.25   & 0.06\\
        \YTVO & 0.65 & 0.45&  0.04\\
        \hline
    \end{tabular}
    \caption{
    Dimensions of the three samples investigated here.
    $L$ = length between contacts; $w$ = width of the sample; $t$ = thickness of the sample.}
    \label{tab:geo_fact}
\end{table}

\begin{figure*}[t]
    \centering
    \includegraphics[width=0.7\textwidth]{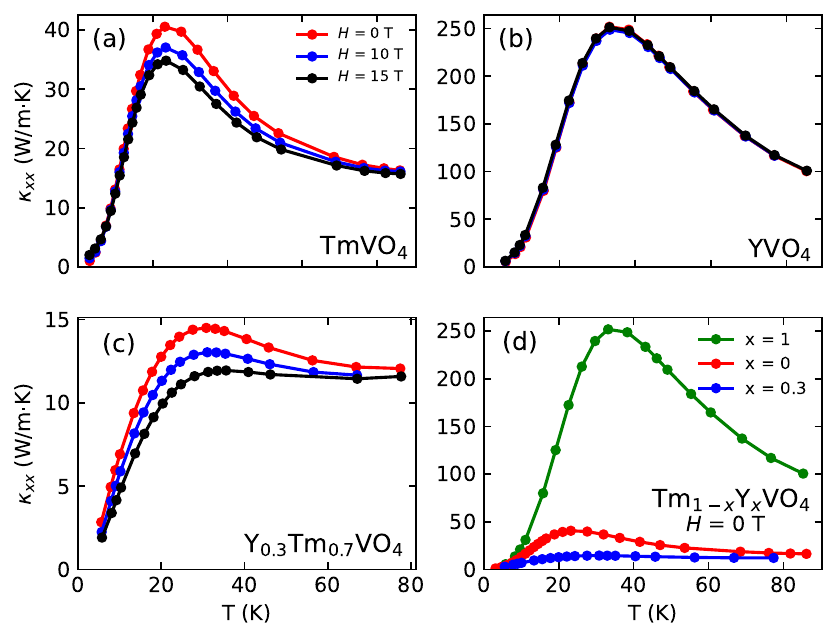}
    \caption{
Thermal conductivity of (a)~\TVO, (b)~\YVO, and~(c)~\YTVO~as a function of temperature, measured with a heat current $J // c$ and 
a magnetic field $H // a$ ($H \perp J$, for $H = 0$ (red), 10~T (blue), and 15~T (black). 
(d)
Comparison of the three conductivities at $H = 0$.
We see that substituting Y for Tm causes a major reduction in $\kappa_{xx}$,
which is then further reduced if Y impurities are added.
%large increase in $\kappa_{xx}$ is observed when Y$^{3+}$ ions are added.
}
\label{fig:Kxx_all_H_4_pannels}
\end{figure*}

%Note that there is a slight misalignment of about 7 \% of the applied H, thus there is about 1 T along the c axis when a field of 15 T is applied out-of-plane along the a axis. 
%
Temperature steps from $\simeq 3$~K to 80 K were done at fixed fields and all the temperature signals were stable before measuring each temperature difference ($\Delta T_{x}$ and $\Delta T_{y}$ in Fig.~\ref{fig:setup}) using type-E (chromium/constantan) thermocouples. Note that a test has been done previously to show that thermocouples and chip-based thermometers (Cernox) yield the same thermal Hall data (see Appendix C in ref. \cite{Gael_thermocouple_cernox}).
Also, note that for all the voltages, the background voltage (when the heat is off) is subtracted from the measured signal (when the heat is on).

The conductivities $\kappa_{xx}$~and $\kappa_{xy}$~were measured as described elsewhere~\cite{2019_Giant_Kxy_pseudogap_cuprates, Grissonnanche_2020_Chiral_phonons, Boulanger_2020_Kxy_NCO_SCOC}. Note that the transverse temperature difference $\Delta T_{y}$ is obtained by measuring in both field polarities ($+ H$ and $- H$). We checked that our technique is independent of the particular choice of sequence, i.e., whether we measure $+ H$ first and then $- H$ or if we measure instead $- H$ first and then $+ H$. This means that our experimental setup and our procedure do not generate any (technique dependent) hysteresis. Then, we antisymmetrize the measured signals to get rid of any longitudinal contribution due to a possible slight misalignment of the transverse contacts:
\begin{equation}
    \Delta T_{y}(T, H)  = [\Delta T_{y}(T, +H) - \Delta T_{y}(T, -H)]/2.
\end{equation}

The sign of $\Delta T_{y}$ is determined by measuring a sample of known $\kappa_{xy}$ with the same setup and wiring. In a metallic sample with Hall coefficient $R_{H} >$ 0 (thus an electrical Hall conductivity $\sigma_{xy} > 0$), the thermal Hall conductivity $\kappa_{xy}$ will be positive (at least in the $T$ = 0 limit) since both thermal and electrical conductivities are related by the Wiedemann-Franz law : $\kappa_{xy}/T = L_{0} \sigma_{xy}$ for $ T \rightarrow$ 0. \\

The thermal conductivity is defined as: 
\begin{equation}
    \kappa_{xx} = \frac{J}{\Delta T_{x} \cdot \alpha},
\end{equation}

where $J$ is the heat current (in W) and $\alpha$ is a geometric factor ($\alpha = \frac{w \cdot t}{L}$; see Table \ref{tab:geo_fact} for sample dimensions). 
The thermal Hall conductivity is defined as: 
\begin{equation}
    \kappa_{xy} = \kappa_{yy} \left (\frac{\Delta T_{y}}{\Delta T_{x}}\right )\left (\frac{L}{w}\right).
\end{equation}

Note that here we assume that $\kappa_{yy} = \kappa_{xx}$ (i.e., that $\kappa_{a} = \kappa_{c}$), but this is not quite right since the $c$ axis conductivity is not identical to the $a$ axis conductivity. 
This implies that the amplitude of $\kappa_{xy}$~reported here is not quite accurate, and would need to be multiplied by the anisotropy factor $\kappa_{a} / \kappa_{c}$. 
Given the needle shape of our samples,  measurements with $J // a$ 
%and $B$ // c axis are quite hard to do and 
have not been done.

\begin{figure*}[t]
    \centering
    \includegraphics[width=1\textwidth]{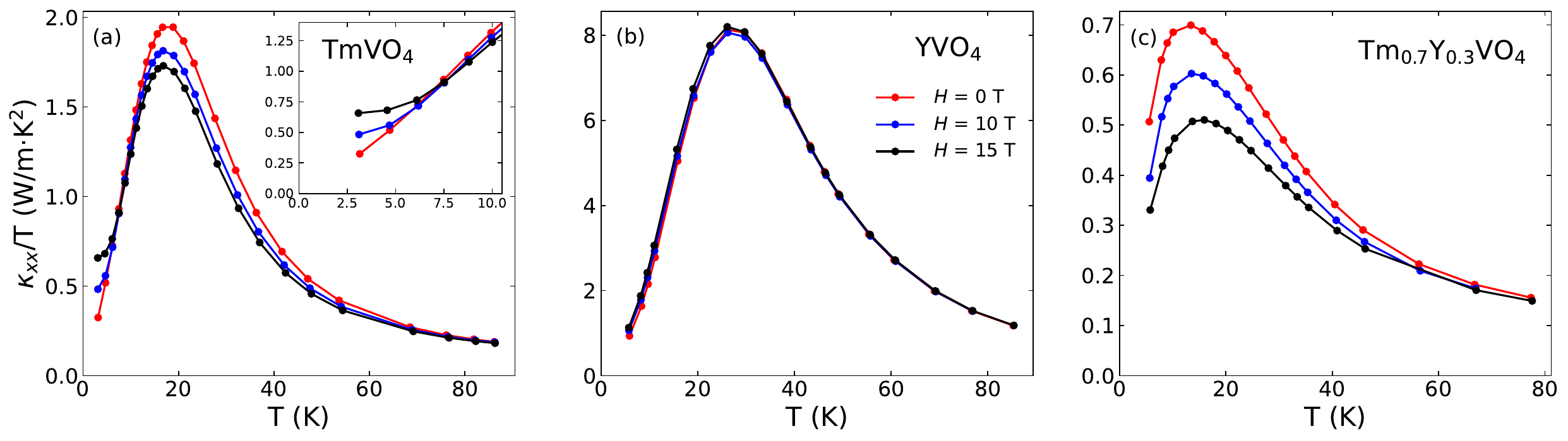}
    \caption{
 Thermal conductivity of (a)~\TVO, (b)~\YVO, and~(c)~\YTVO, for different applied magnetic fields, as indicated,
plotted as $\kappa_{xx}$/$T$ versus $T$ to emphasize the low-temperature regime.
Note the strong field dependence in the two samples that contain Tm$^{3+}$ ions,
compared to the much weaker field dependence in \YVO.
In panel (a), the inset shows a zoom below 10~K, where we see a different regime of $H$ dependence,
with a strong increase in $\kappa_{xx}$~with increasing $H$.}
    \label{fig:Kxx_T_comp_15T_3panels}
\end{figure*}

\section{RESULTS}

\subsection{Thermal conductivity, $\kappa_{xx}$}
The thermal conductivity $\kappa_{xx}$~of our three samples is displayed in Fig.~\ref{fig:Kxx_all_H_4_pannels}.
All three curves are typical of insulators, with a peak near $T \simeq$ 20-30 K.

The thermal conductivity of \TVO~was measured previously~\cite{Daudin_1982}. 
These early
measurements were done with $J // H // c$ (along the easy axis \cite{Suzuki_Inoue_Ohtsuka_1981}) for $H < 7$~T and for $T < 30$~K,
whereas
our measurements were done with $J // c$ and $H // a$,
covering a larger temperature and field range. 
This previous study reports a 
slightly
higher $\kappa_{xx}$ peak,
perhaps
due to larger samples and higher sample quality. 
The field was found to increase $\kappa_{xx}$ at low temperatures ($T<~6$~K). In our study, 
we also see an
 increase in $\kappa_{xx}$ with $H$ at low temperatures [$T < 10$~K; inset of Fig.~\ref{fig:Kxx_T_comp_15T_3panels}(a)].

The zero-field curves are compared in Fig.~\ref{fig:Kxx_all_H_4_pannels}(d),
where we see that $\kappa_{xx}$~in \YVO~is much larger than in the other two samples.
The magnitude of $\kappa_{xx}$~in the stoichiometric sample of paramagnetic \TVO~is much lower presumably because phonons scatter on the crystal field levels of the Tm$^{3+}$ ions.
This is also presumed to be the case in the frustrated spin system Tb$_2$Ti$_2$O$_7$, where the scattering of phonons would involve the crystal field levels of the Tb$^{3+}$ ions, as argued previously \cite{Mori_2014}. This scattering process is much stronger for Tb$^{3+}$ than Tm$^{3+}$, resulting in a value of $\kappa_{xx}$~$\simeq 2~$W/m$\cdot$K at $T = 20$~K in Tb$_2$Ti$_2$O$_7$ \cite{Sun_Pyro_Kxx},
compared to $\kappa_{xx}$~$\simeq 40$~W/m$\cdot$K in \TVO~[Fig.~\ref{fig:Kxx_all_H_4_pannels}(a)].

\begin{figure}[b]
    \centering
    \includegraphics[width=0.49\textwidth]{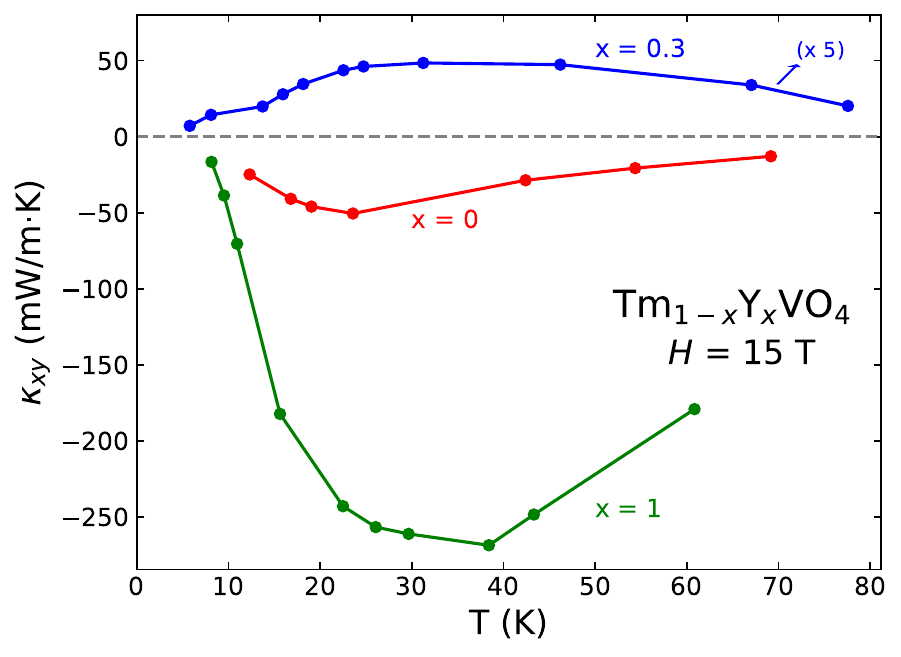}
    \caption{
Thermal Hall conductivity of 
\TVO~(red), 
\YVO~(green), 
and 
\YTVO~(blue)
as a function of temperature, measured with a heat current $J // c$ and 
a magnetic field $H // a$ ($H \perp J$), at $H = 15$~T.
%
%for $H$ = 15 T with $J$ // c and $H$ // a. \TVO \hspace{0.1 em} and \YVO \hspace{0.1 em} have a negative $\kappa_{xy}$ while \YTVO \hspace{0.1 em} has a positive one. 
%
Note that $\kappa_{xy}$ for \YTVO \hspace{0.1 em} has been multiplied by a factor of 5. 
%
%\TVO, \YVO \hspace{0.1 em} and \YTVO\hspace{0.1 em} peak at the same temperature as their corresponding $\kappa_{xx}$ data, i.e., at 25 K, 35 K and 30 K respectively.
}
    \label{fig:Kxy_15T_all}
\end{figure}

As in the case of Tb$_2$Ti$_2$O$_7$, where the field dependence of $\kappa_{xx}$~is exceptionally strong~\cite{Sun_Pyro_Kxx},
we view the sizable field dependence of $\kappa_{xx}$ in \TVO~[Fig.~3(a)] and  \YTVO~[Fig.~\ref{fig:Kxx_all_H_4_pannels}(c)],
as a confirmation that phonons are scattered by the crystal field levels of the Tm$^{3+}$ ions.
The weaker field dependence seen in \YVO~[Fig.~\ref{fig:Kxx_all_H_4_pannels}(b)] is consistent with that picture.
We attribute the fact that $\kappa_{xx}$~in \YTVO~is even smaller than in \TVO~to the extra disorder introduced by the Y impurities [Fig.~\ref{fig:Kxx_all_H_4_pannels}(d)].

In Fig.~\ref{fig:Kxx_T_comp_15T_3panels}, we plot $\kappa_{xx}$$/T$ versus $T$, to emphasize the data at low temperature.
We see that below $\sim$ 10~K or so, the field dependence of $\kappa_{xx}$~in \TVO~becomes very strong and opposite in sign
relative to its behavior at higher temperature [inset of Fig.~\ref{fig:Kxx_T_comp_15T_3panels}(a)]. We speculate that this change of behavior is associated with the quadratic splitting of the crystal field levels of the Tm$^{3+}$ ions with field (second-order Zeeman interaction for $H$//a) \cite{vinograd_shirer_massat_2022} which changes the resonant phonon scattering (for phonons of appropriate symmetry). 

Interestingly, this change in field dependence, as the temperature is raised, was actually captured by theoretical calculations that consider the main scattering mechanism of phonons to be resonant scattering on electronic levels of the Tm$^{3+}$ ions~\cite{Mutscheller_1986,Mutscheller_1987}, albeit for temperatures much closer to the transition.

\subsection{Thermal Hall conductivity, $\kappa_{xy}$}
The thermal Hall conductivity $\kappa_{xy}$~of our three samples is displayed in Fig.~\ref{fig:Kxy_15T_all}.
Only data at $H = 15$~T are shown, but data at $H = 10$~T were also taken;
they are similar but smaller in magnitude, roughly in proportion to the field amplitude.
The first observation is that all three samples exhibit a non-negligible thermal Hall conductivity.
The surprise is that the sign of $\kappa_{xy}$~is negative in the two stoichiometric materials, 
\TVO~and \YVO,
but it is positive in the disordered sample \YTVO.

We see that the temperature at which $\kappa_{xy}$$(T)$ peaks (Fig.~\ref{fig:Kxy_15T_all}) is roughly the same as the temperature at which the phonon-dominated
$\kappa_{xx}$$(T)$ peaks (Fig.~\ref{fig:Kxx_all_H_4_pannels}), for example, $\simeq 25$~K in \TVO~and $\simeq 35$~K in \YVO.
As argued before for other materials~\cite{Sr2TiO4_Kamran_2020,Boulanger_2020_Kxy_NCO_SCOC,CTO_Lu_2022,Kamran_2023_Black_phosphorous}, this is an argument in support of phonons being the heat carriers responsible for the thermal Hall signal
in these materials.

%$\kappa_{xy}$ measurements from $T$ = 5 K to $T$ = 80 K with $H$ = 10 and 15 T were done on the same three materials. In Fig.~\ref{fig:Kxy_15T_all}, only the $H$ = 15 T data is shown. A negative $\kappa_{xy}$ is measured in \TVO \hspace{0.1em} and \YVO \hspace{0.1em}. However, $\kappa_{xy}$ for \YTVO \hspace{0.1em} is positive. All $\kappa_{xy}$ curves have the same peak $T$ as in $\kappa_{xx}$($T$) for each material. \YVO \hspace{0.1em} has the largest $\kappa_{xy}$ peak value. Note that for $H$ = 10 T, a similar conclusion is drawn, but the magnitude of $\kappa_{xy}$ is slightly smaller than for $H$ = 15 T.

%When looking at the degree of chirality corresponding to the ratio of $\kappa_{xy}$ over $\kappa_{xx}$, it is observed that at $H$ = 15 T and $T$ = 20 K, the ratio is around 1 x 10$^{-3}$. This value is typical for insulators \cite{CTO_Lu_2022}.

Looking at Fig.~\ref{fig:Kxy_15T_all},
we also notice that the magnitude of $\kappa_{xy}$~appears to roughly scale with the magnitude of $\kappa_{xx}$~seen in Fig.~\ref{fig:Kxx_all_H_4_pannels}(d).
This is confirmed when looking at the thermal Hall angle, plotted as $|\kappa_{xy} / \kappa_{xx}|$ versus $T$ in Fig.~\ref{fig:Kxy_Kxx_15T_all}.
Indeed, we see that $|\kappa_{xy} / \kappa_{xx}| \simeq 1 \times 10^{-3}$ at $T \sim 20$~K in both \YVO~and \YTVO,
even though the magnitude of $\kappa_{xy}$~is 25 times larger in \YVO~(Fig.~\ref{fig:Kxy_15T_all}).
This is because $\kappa_{xx}$~is also roughly 25 times larger in \YVO~(Fig.~\ref{fig:Kxx_all_H_4_pannels}).

\section{DISCUSSION}

\begin{figure}[t]
    \centering
    \includegraphics[width=0.46\textwidth]{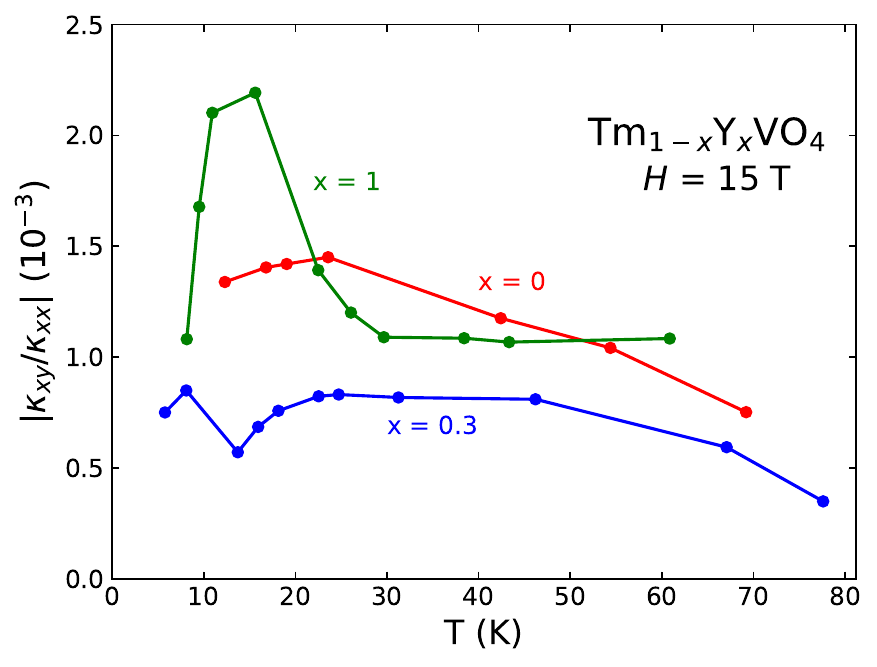}
    \caption{
Thermal Hall angle of \TVO~(red), \YVO~(green), and \YTVO~(blue), plotted as $|$$\kappa_{xy}$/$\kappa_{xx}$$|$ versus $T$, for $H = 15$~T.
At $T = 20$~K, the obtained value is around 1 $\times$ 10$^{-3}$ for all three materials, a magnitude typical of the phonon Hall effect in several insulators~\cite{CTO_Lu_2022,Kamran_2023_Black_phosphorous} (see Fig. \ref{fig:ratio_pthe}).
%Note that for $T \geq$ 20 K, the Hall angle is close to 1 x 10$^{-3}$ for all three materials.
}
    \label{fig:Kxy_Kxx_15T_all}
\end{figure}

%\subsubsection{Field dependence}
%All three materials exhibit a field-dependence. 
%At high temperatures, $\kappa_{xx}$ of \TVO \hspace{0.1 em} decreases with increasing $H$ especially for $T >$ 8 K (Fig. \ref{fig:Kxx_T_comp_15T_3panels} (a)). When Tm is fully replace with Y, the field dependence becomes much smaller (Fig. \ref{fig:Kxx_T_comp_15T_3panels} (c)). It is around 2 \% for $T >$ 25 K. As 30 \% of Y is added in \TVO, $\kappa_{xx}$ decreases with increasing $H$ (Fig. \ref{fig:Kxx_T_comp_15T_3panels} (c)). The field dependence ranges from 20 \% to 5 \% for $T >$ 30 K. \\

%At low temperatures, $\kappa_{xx}$ of \TVO \hspace{0.1 em} actually increases with increasing $H$ (for $T <$ 8 K, see inset in Fig. \ref{fig:Kxx_T_comp_15T_3panels} (a)). This low $T$ behaviour is believed to maybe be related to the cooperative Jahn-Teller and the ferroquadrupolar transitions at $T_{D} \sim$ 2 K. Note that for the 30 \% Y-substituted material, no more low $T$ behaviour is observed and a field dependence of about 35 \% is observed at about 10 K. \\

%\subsection{Thermal Hall conductivity, $\kappa_{xy}$}

%\subsubsection{Effect of Tm}

In the pyrochlore study, 
a non-zero (and positive) thermal Hall effect was measured in Tb$_{2}$Ti$_{2}$O$_{7}$,
but a negligible $\kappa_{xy}$~was detected in Y$_{2}$Ti$_{2}$O$_{7}$ \cite{Hirschberger_2015_pyrochlores}. 
It was initially speculated that the thermal Hall signal in Tb$_{2}$Ti$_{2}$O$_{7}$ was due to some exotic spin excitations \cite{Hirschberger_2015_pyrochlores}. 
However, a subsequent study found that the $\kappa_{xy}$~signal remains as strong when 70\% of the magnetic Tb$^{3+}$ ions are replaced by nonmagnetic Y$^{3+}$ ions,
thereby dramatically altering the frustrated spin-liquid state of pure Tb$_{2}$Ti$_{2}$O$_{7}$~\cite{Phononic_thermal_Hall_effect_in_diluted_terbium_oxides}.
This lead to the conclusion that phonons are, in  fact, the heat carriers responsible for the thermal Hall effect in these pyrochlores.
Moreover,
the fact that $\kappa_{xy}$~becomes negligible in Y$_{2}$Ti$_{2}$O$_{7}$, when all Tb is replaced by Y,
suggests that magnetism plays a key role in the generation of the phonon thermal Hall effect in these pyrochlores.

%that measured an intermediate Y concentration ([Tb$_{0.3}$Y$_{0.7}$]$_{2}$Ti$_{2}$O$_{7}$) found an increase in $\kappa_{xy}$ compared to the almost 0 thermal Hall conductivity measured in Y$_{2}$Ti$_{2}$O$_{7}$. This directly contradicts the spin excitation picture and actually attributes a phononic origin to $\kappa_{xy}$ in these pyrochlores~\cite{Phononic_thermal_Hall_effect_in_diluted_terbium_oxides}.

Turning to our own study,
we also observe a non-zero $\kappa_{xy}$~in \TVO, another oxide with magnetic ions (Tm$^{3+}$),
and we also attribute this Hall effect to phonons, as argued above.
In our comparison with Tb$_{2}$Ti$_{2}$O$_{7}$,
we see two differences.
The first is the fact that our nonmagnetic parent compound, \YVO,
also displays a non-zero $\kappa_{xy}$, unlike the negligible signal of Y$_{2}$Ti$_{2}$O$_{7}$.
So in these vanadates, the phonon Hall effect does not
depend crucially on the magnetic moment associated with Tm$^{3+}$ ions.
Note that a non-zero thermal Hall effect from phonons has been observed in other 
nonmagnetic insulators, such as strontium titanate~\cite{Sr2TiO4_Kamran_2020} and black phosphorous~\cite{Kamran_2023_Black_phosphorous}.

The second difference is the sign of $\kappa_{xy}$: 
positive in Tb$_{2}$Ti$_{2}$O$_{7}$, negative in \TVO.
The sign of the phonon Hall conductivity is an entirely open question, on which existing theories shed little light.
Phenomenologically, both signs are observed, see Fig. \ref{fig:ratio_pthe}.
For example, the phononic $\kappa_{xy}$~is negative in 
SrTiO$_{3}$~\cite{Sr2TiO4_Kamran_2020}, 
Cu$_{3}$TeO$_{6}$~\cite{CTO_Lu_2022}, and
black phosphorus~\cite{Kamran_2023_Black_phosphorous},
as well as in all cuprates, 
whether undoped~\cite{Grissonnanche_2020_Chiral_phonons,Boulanger_2020_Kxy_NCO_SCOC}, 
hole-doped~\cite{2019_Giant_Kxy_pseudogap_cuprates},
or
electron-doped~\cite{Boulanger_2022_Kxy_electron_dopes_cuprates}.
A positive phononic $\kappa_{xy}$~is observed in
Fe$_{2}$Mo$_{3}$O$_{8}$ \cite{THE_Multiferroics}, 
(Tb$_{0.3}$Y$_{0.7}$)$_{2}$Ti$_{2}$O$_{7}$~\cite{Phononic_thermal_Hall_effect_in_diluted_terbium_oxides},
and 
RuCl$_{3}~$\cite{RuCL3_Etienne_2022,Hentrich_PRB_2019}.
A remarkable feature of our findings is that $\kappa_{xy}$~changes sign to positive upon introducing 30\% Y into \TVO~(Fig.~\ref{fig:Kxy_15T_all}).
This suggests that impurities, in much larger concentration in \YTVO~relative to either \TVO~or \YVO,
is an important ingredient for the generation of a phonon thermal Hall effect.

It is instructive to look at the magnitude of $\kappa_{xy}$.
As argued by others~\cite{THE_Multiferroics,CTO_Lu_2022,Kamran_2023_Black_phosphorous}, 
the relevant quantity for this is the ratio $\kappa_{xy}$/$\kappa_{xx}$, namely the Hall angle.
In Fig.~\ref{fig:Kxy_Kxx_15T_all},
we see that in our three samples $|$$\kappa_{xy}$/$\kappa_{xx}$$| \simeq 1 \times 10^{-3}$ at $T = 20$~K and $H = 15$~T.
This is quite typical.
Indeed,
the cuprate Mott insulators La$_{2}$CuO$_{4}$, Nd$_{2}$CuO$_{4}$, and Sr$_{2}$CuO$_{2}$Cl$_{2}$ and the antiferromagnetic insulator Cu$_{3}$TeO$_{6}$ all 
have a Hall angle of about $3 \times 10^{-3}$ at the same temperature and field~\cite{CTO_Lu_2022}.
The Kitaev spin liquid candidate RuCl$_{3}$ has a Hall angle of about $1 \times 10^{-3}$.
Note that the magnitude of $\kappa_{xy}$~varies by three orders of magnitude across those various materials (see Fig. \ref{fig:ratio_pthe}).

It is worth noting that significantly larger thermal Hall angles have been observed in two cases so far:
the Ce-doped cuprate Nd$_2$CuO$_4$~\cite{Boulanger_2022_Kxy_electron_dopes_cuprates} 
and the Rh-doped iridate Sr$_2$IrO$_4$~\cite{ataei2023impurityinduced}.
In as-grown samples of Nd$_2$CuO$_4$ with 11\% Ce
and Sr$_2$IrO$_4$ with 5\% Rh, 
$|$$\kappa_{xy}$/$\kappa_{xx}$$| \simeq 30 \times 10^{-3}$ at $T = 20$~K and $H = 15$~T.
Both materials are insulators with antiferromagnetic order at these dopings,
and it was argued in the latter study that impurities embedded in an antiferromagnetic environment
strongly promote the phonon thermal Hall effect~\cite{ataei2023impurityinduced},
as suggested theoretically for a mechanism of resonant side-jump scattering of phonons~\cite{Guo_Joshi_Sachdev_2022}.

\section{SUMMARY}

We measured a non-zero thermal Hall conductivity $\kappa_{xy}$~in the vanadates 
\TVO, \YVO~and 
\YTVO.
All evidence points to phonons as the heat carriers responsible for generating this Hall effect.
The magnitude of the Hall response is similar in all three, with a Hall angle of
$|\kappa_{xy}$/$\kappa_{xx}| \simeq 1 \times 10^{-3}$ at $T = 20$~K and $H = 15$~T, 
comparable to the phonon Hall effect in several insulating materials.
This shows that Tm$^{3+}$ ions do not play an essential role in generating the Hall effect in these rare-earth vanadates. 
While the sign of $\kappa_{xy}$~in the two stoichiometric compounds \TVO~and \YVO~is negative,
we find a positive sign in the more disordered \YTVO. 
This points to a special role played by impurities in this family of materials.

%A non-zero and negative $\kappa_{xy}$ was obtained for \YVO \hspace{0.1 em} as well as for \TVO \hspace{0.1 em}. Surprisingly, the 30 \% Y-substituted compound also has a non-zero $\kappa_{xy}$, but with a positive sign. Note that the sign of the signal itself is still not understood for insulators. However, an additional mechanism involving impurities might need to be considered to explain why there is such a sign change. 

%Despite the sign and magnitude of the measured conductivities, the Hall angle remains comparable for all three vanadates ($|\kappa_{xy}$/$\kappa_{xx}|$ = 1 x 10$^{-3}$). 

%\textcolor{red}{\bf LOUIS STOPPED HERE}

%% ///////////////////////////////////////////////////////////////////////////////////////////////////////////////////////////////////////////////////////////////////////////

\section*{ACKNOWLEDGMENTS}

We thank S. Fortier for his assistance with the experiments and A. Ataei, P. Fournier and I. Garate for stimulating discussions. 
L. T. acknowledges support from the Canadian Institute for Advanced Research (CIFAR) as a CIFAR Fellow 
and funding from the Institut Quantique, the Natural Sciences and Engineering Research Council of Canada (NSERC; PIN:123817), 
the Fonds de Recherche du Qu\'ebec - Nature et Technologies (FRQNT), the Canada Foundation for Innovation (CFI), 
and a Canada Research Chair. 
This research was undertaken thanks in part to funding from the Canada First Research Excellence Fund.

Crystal growth and characterization performed at Stanford University was supported by the Air Force Office of Scientific Research under Award No. FA9550-20-1-0252. MPZ was also partially supported by a National Science Foundation Graduate Research Fellowship under Grant No. DGE-1656518.

\section*{APPENDIX}

The thermal Hall effect can be measured in three ways and we have tested these methods on the hole-doped cuprate La$_{2-x}$Sr$_{x}$CuO$_{4}$ $p$ = 0.06 (LSCO).\\

\begin{figure}[h]
    \centering
    \includegraphics[width=0.5\textwidth]{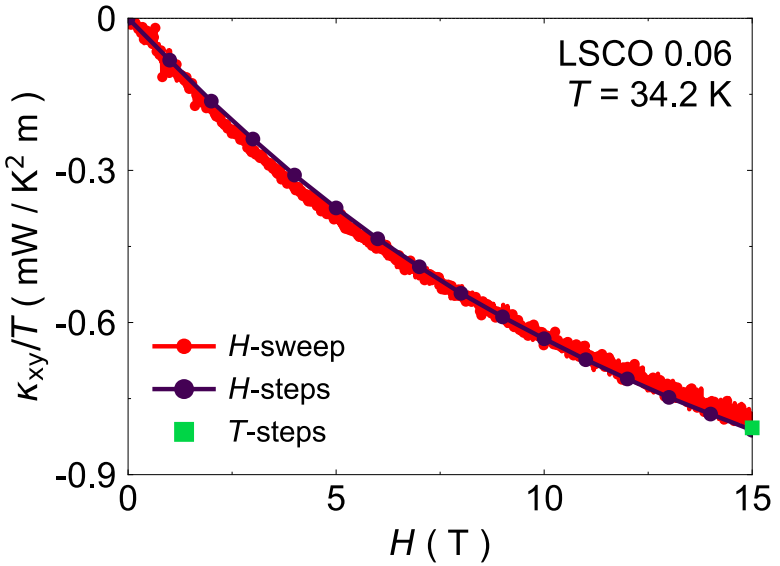}
    \caption{$\kappa_{xy}/T$ as a function of magnetic field $H$ at fixed $T$ = 34.2 K in the hole-doped cuprate La$_{2-x}$Sr$_{x}$CuO$_{4}$ $p$ = 0.06 (LSCO). The red circles are obtained sweeping the magnetic field continuously at a rate of 0.7 T/min. The purple circles are obtained using a steady-state method as a function of field with steps of 1 T. The green square is obtained using a steady-state method as a function of temperature at $H$ = 15 T.}
    \label{fig:Three_techs}
\end{figure}

\begin{figure}[h]
    \centering
    \includegraphics[width=0.5\textwidth]{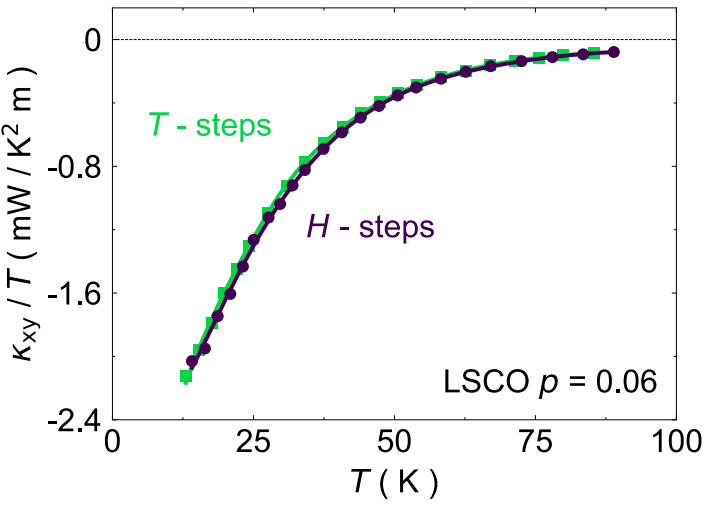}
    \caption{$\kappa_{xy}/T$ as a function of temperature at $H$ = 15 T in the hole-doped cuprate La$_{2-x}$Sr$_{x}$CuO$_{4}$ $p$ = 0.06 (LSCO). The green squares are obtained using a steady-state method as a function of temperature. The purple circles are obtained using a steady-state method as a function of magnetic field.}
    \label{fig:steps_techs}
\end{figure}

(1) Steady-state method as function of temperature, also known as ``$T$-steps" (see Figs. \ref{fig:Three_techs} and \ref{fig:steps_techs}). The magnetic field $H$ is kept fixed (at $+H$) while the temperature $T$ of the sample is changed in discrete steps (2 to 3 K steps). At each temperature, the background voltage of the thermocouple that measures $\Delta T_{y}$ is recorded before the heat is applied to the sample. When the sample is in an equilibrium configuration, $\Delta T_{y} (H)$ is measured. 

The background voltage of the thermocouple is carefully subtracted from the measured heat-on signal from the thermocouple. Once the entire temperature range is covered, say from 10 K to 100 K, the field direction is reversed to -$H$. The same procedure is followed for this field polarity. Then, the $\Delta T_{y}$ signal is antisymmetrized; $\Delta T_{y}(T, H)  = [\Delta T_{y}(T, +H) - \Delta T_{y}(T, -H)]/2$, such that any symmetric contribution, coming from the longitudinal thermal gradient due to misalignment of the transverse contacts, is removed. \\

(2) Steady-state method as a function of field, also known as ``$H$-steps'' (see Figs. \ref{fig:Three_techs} and \ref{fig:steps_techs}). The temperature $T$ is kept fixed, and the magnetic field $H$ is changed in discrete steps (1 to 2 T steps). At a fixed temperature, a heat current $\vec{J}$ is sent to the sample, and once it reaches equilibrium, the magnetic field is changed from $+H$ to -$H$ in steps. For each field step $h$, we define $\Delta T_{y}(T,h)  = [\Delta T_{y}(T, +h) - \Delta T_{y}(T, -h)]/2$. Once the entire magnetic field range is covered, the temperature of the sample is changed and the same procedure is repeated. \\ 

(3) Field sweep method, also known as ``$H$-sweeps'' (see Fig. \ref{fig:Three_techs}). This method is similar to 2), but now the magnetic-field is continuously changed from $+H$ to -$H$.

We found satisfactory agreement between methods (1), (2), and (3), see Fig. \ref{fig:Three_techs}.

%
%% ///////////////////////////////////////////////////////////////////////////////////////////////////////////////////////////////////////////////////////////////////////////

%\clearpage
\bibliography{reference}

\end{document}